\documentclass[conference]{ieeeconf}
\IEEEoverridecommandlockouts
\usepackage{amsmath,amssymb,amsfonts}
\usepackage{mathrsfs}
\usepackage{algorithmic}
\usepackage{graphicx}
\usepackage{textcomp}
\usepackage{hyperref}
\usepackage{lipsum}
\usepackage[ruled,vlined,linesnumbered]{algorithm2e}
\usepackage{cleveref}
\usepackage{subcaption}
\usepackage{tikz}
\usepackage{float}
\usepackage[inkscapelatex=false]{svg}
\usepackage{paralist}
\usepackage{xcolor}

\usepackage[bibstyle=ieee, citestyle=numeric-comp, doi=false,isbn=false,url=false,eprint=false, backend=biber, sorting=none]{biblatex}

\usetikzlibrary{positioning}

\newtheorem{problem}{Problem}[section]
\newtheorem{remark}{Remark}[section]

\crefname{algocf}{algorithm}{algorithms}
\Crefname{algocf}{Algorithm}{Algorithms}
\def\BibTeX{{\rm B\kern-.05em{\sc i\kern-.025em b}\kern-.08em
    T\kern-.1667em\lower.7ex\hbox{E}\kern-.125emX}}
\begin{document}

\definecolor{lightpurple}{rgb}{0.93,0.88,0.97}
\definecolor{darkpurple}{rgb}{0.42,0.20,0.60}
\definecolor{wheat}{rgb}{0.96,0.87,0.70}
\newcommand{\todo}[1]{
    \begin{center}
        \setlength{\fboxrule}{1.2pt}%
        \fcolorbox{darkpurple}{lightpurple}{\parbox[t]{0.45\textwidth}{\textbf{\textcolor{darkpurple}{ToDo:}} #1}}
    \end{center}}

\newcommand{\kevin}[1]{
    \begin{center}
        \setlength{\fboxrule}{1.2pt}%
        \fcolorbox{darkpurple}{wheat}{\parbox[t]{0.45\textwidth}{\textbf{\textcolor{darkpurple}{Kevin:}} #1}}
    \end{center}}

\newcommand{\robustnessFunc}[2]{\rho(#1, #2)}
\newcommand{\nodeLabelFunc}{\mathcal{L}_\text{node}}
\newcommand{\nodeLabelSet}{{\Pi_\text{node}}}
\newcommand{\nodeLabelElement}{{\pi_\text{node}}}
\newcommand{\nodeSet}{V}
\newcommand{\nodeElement}{v}
\newcommand{\edgeSet}{E}
\newcommand{\signalSet}{S}
\newcommand{\signalElement}{s}
\newcommand{\signalLabelSet}{{\Pi_\text{signal}}}
\newcommand{\signalLabelElement}{{\pi_\text{signal}}}
\newcommand{\signalLabelFunc}{{\mathcal{L}_\text{signal}}}
\newcommand{\nodeToInt}{\iota}
\newcommand{\timeStep}{k}

\newcommand{\always}[2][a,b]{G_{[#1]}#2}
\newcommand{\eventually}[2][a,b]{F_{[#1]}#2}
\newcommand{\until}[3][a,b]{#2 \mathcal{U}_{[#1]} #3}

\title{Satisfiability Modulo Theories for Robust Synthesis of Signal Temporal Logic Specifications}

\author{
Philip Smith\textsuperscript{1}, Kevin Leahy\textsuperscript{1}
\thanks{$^1$ Department of Robotics Engineering, Worcester Polytechnic Institute, {\tt\small \{psmith3,kleahy\}@wpi.edu}}
\thanks{DISTRIBUTION STATEMENT A. Approved for public release; distribution is unlimited. OPSEC \#11197}
}

\maketitle

\begin{abstract}
Signal Temporal Logic is a popular framework for describing a robotic system's goals. Once described using a temporal logic framework, synthesis turns the goals into actual steps for the robots to follow. Currently, the state of the art for sound and complete synthesis is using Mixed-Integer Linear Programming (MILP), which struggles on long time horizons and large environments, preventing the construction of guaranteed safe plans for long deployments. This paper introduces a new synthesis method using Satisfiability Modulo Theories (SMT) which scales better for long deployments and adapts to environmental changes more quickly. We do this by introducing an encoding method for STL into SMT constraints, and an encoding method for discrete-time dynamical systems modeled as a transition system. We then conduct an empirical comparison of our SMT based methods against the state of the art MILP based methods. Results show that for this class of problems, SMT is an effective encoding strategy that performs the best on long time horizon tasks.
\end{abstract}

\section{Introduction}\label{sec:intro}


Temporal logics are a popular approach for precisely specifying complex tasks and behaviors for cyber-physical systems~\cite{konurSurveyTemporalLogics2013, baierPrinciplesModelChecking2008}. Signal Temporal Logic (STL) is one such logic, defined over real-valued signals, making it a natural fit for cyber-physical systems~\cite{malerMonitoringTemporalProperties2004}. STL has formally defined quantitative semantics, often called robustness ($\rho$), which indicates not just if a signal satisfies a specification, but by how much~\cite {donzeRobustSatisfactionTemporal2010, fainekosRobustnessTemporalLogic2009}.  Synthesis is the problem of automatically constructing a system (often a control sequence) that satisfies a specification, rather than designing something and checking it after the fact~\cite{beltaFormalMethodsControl2019b}. If synthesis is fast enough, it can be included as part of the control loop\cite{ramanModelPredictiveControl2017}, ensuring that a system handles disturbances smoothly and without violating the specifications, for example, by replanning in the event of an unexpected infeasibility.

The dominant approach to STL synthesis today is Mixed-Integer Linear Programming, which encodes the Boolean structure of an STL formula using binary variables~\cite{ramanModelPredictiveControl2014}. Despite improvements to their encoding~\cite{kurtzMixedIntegerProgrammingSignal2022, kurtzBayesianOptimizationPolynomial2019}, MILP-based solvers scale poorly; every additional operator or nesting level adds more binary variables, and MILP solvers slow down fast as that count grows~\cite{kurtzMixedIntegerProgrammingSignal2022}. MILP complexity is exponential in both the time bound of the formula and the number of predicates~\cite{kurtzBayesianOptimizationPolynomial2019}. In practice, this caps the size and complexity of specifications that can be synthesized. 

Many other approaches have been researched for STL synthesis. Counterexample-guided inductive synthesis (GEGIS) methods iteratively refine a candidate signal against a falsifying environment, avoiding the upfront combinatorial encoding but sacrificing the completeness guaranteed by MILP-based methods~\cite{ramanReactiveSynthesisSignal2015, ramanModelPredictiveControl2017}. Sampling-based and automaton-guided search methods also trade completeness, only giving a probabilistic guarantee. While they scale well in practice, they do not certify optimality~\cite {hoAutomatonGuidedControlSynthesis2022}. Another line of work reformulates the STL robustness function into a smooth differentiable objective which enables gradient-based trajectory optimization~\cite {gilpinSmoothRobustnessMeasure2020}. These methods are fast and effective for local optima but give up soundness and global optimality guarantees~\cite {hanExactSmoothReformulations2025}. Neural approaches learn a mapping from specification to control signal directly, which reduces the runtime cost by spending it on training data and offers no correctness guarantees on unseen or out-of-distribution specifications~\cite{hashimotoNeuralControllerSynthesis2022}.

Satisfiability Modulo Theories (SMT) is an extension of the Boolean satisfiability problem (SAT) adding in theories, such as the theory of Reals~\cite{demouraModelbasedTheoryCombination2008}. Optimization Modulo Theories (OMT)\footnote{This paper will refer to OMT as SMT for ease of recognition} extends  SMT with the ability to optimize an objective under the same constraints~\cite{tsiskaridzeGeneralizedOptimizationModulo2024}. While SMT has been used for verification and monitoring of STL formulas before~\cite{malerMonitoringTemporalProperties2004, yuSTLmcRobustSTL2022, roehmSTLInspectorSTLValidation2017, satoOptimizationBasedModelChecking2024, leeEfficientSMTBasedModel2021}, it has not been used for synthesis. We posit that the native capability for solving Boolean logic naturally provides an advantage for SMT in STL synthesis compared to MILP approaches, provided that an efficient dynamics representation can be found. We also posit that the structure of our dynamics for SMT will allow for a more efficient replan utilizing SMT's native replan strength~\cite{aquinoReusingSolutionsModulo2021}. In contrast, this is something MILP struggles with, as branch-and-bound warm start requires well-tuned heuristics and design~\cite{marcucciWarmStartMixedInteger2021}. 

The contributions of this paper are: \begin{inparaenum} \item An OMT-based robustness-maximizing encoding for STL synthesis; \item A discrete transition system encoding for OMT synthesis; \item An empirical comparison on discrete-type dynamics modeled as a transition system, showing efficiency gains and improvements to the replanning capabilities. \end{inparaenum}

\section{Problem Formulation}\label{sec:problem_formulation}
In this work, we consider a task specified using signal temporal logic (STL). STL is a temporal logic with the following grammar~\cite{malerMonitoringTemporalProperties2004}:
\begin{equation} \label{STL_syntax}
    \phi\;::=\;\text{pred}\mid 
    \eventually{\phi} \mid
    \always{\phi} \mid
    \until{\phi_1}{\phi_2} \mid \neg \phi \mid \phi_1\land\phi_2
\end{equation}
Where pred is a predicate relating a real-valued signal variable to a threshold ($\epsilon$) written as $v \geq \epsilon$. In addition, using $\land$ and $\neg$ we can derive a $\lor$ operator. 
\begin{remark}[STL semantics]
    $\eventually{}$ is the “eventually” operator, and specifies that the following formula will eventually be true within the time frame $[a, b]$;
    $\always{}$ is the “always” operator, and specifies that the following formula will hold for the entirety of the time frame $[a,b]$;
    $\until{}{}$ is the “until” operator, and specifies that the first formula will hold until the second formula is true, and that switch will happen sometime in the time frame $ [a,b] $.
\end{remark}

\subsection{Boolean Semantics}
STL Boolean semantics is defined recursively as follows~\cite{beltaFormalMethodsControl2019b}:
\begin{subequations}
\label{eq:boolean_semantics}
\begin{align}
    (s,t) &\models (v \ge \epsilon) \iff v \ge \epsilon \\
    (s,t) &\models (v \le \epsilon) \iff v \le \epsilon \\
    (s,t) &\models \neg \phi \iff (s,t) \not \models \phi \\
    (s,t) &\models \phi_1 \land \phi_2 \iff (s,t) \models \phi_1 \land (s,t) \models \phi_2 \\
    (s,t) &\models \phi_1 \lor \phi_2 \iff (s,t) \models \phi_1 \lor (s,t) \models \phi_2 \\
    (s,t) &\models \eventually{\phi} \iff \exists t^\prime \in [a,b], (s,t^\prime) \models \phi \\
    (s,t) &\models \always{\phi} \iff \forall t^\prime \in [a,b], (s,t^\prime) \models \phi \\
    (s,t) &\models \until{\phi_1}{\phi_2} \iff \exists t^\prime \in [a,b], (s,t^\prime) \models \phi_2\land \notag \\ &\qquad \forall t^{\prime\prime} \in [a,t^\prime], (s,t^{\prime\prime}) \models \phi_1
\end{align}
\end{subequations}
Where $(s,t) \models \phi$ is read as signal suffix $(s,t)$ satisfies formula $\phi$. And $v$ is the value of a signal related to a threshold $\epsilon$.

\subsection{Robustness Semantics}
STL Robustness semantics is defined recursively~\cite{beltaFormalMethodsControl2019b}:
\begin{subequations}
    \label{eq:robustness_semantics}
    \begin{align}
        \robustnessFunc{v \ge \epsilon}{t} &= v - \epsilon \\
        \robustnessFunc{v \le \epsilon}{t} &= \epsilon - v \\
        \robustnessFunc{\neg \phi}{t} &= - \robustnessFunc{\phi}{t} \\
        \robustnessFunc{\phi_1 \land \phi_2}{t} &= \min( \robustnessFunc{\phi_1}{t}, \robustnessFunc{\phi_2}{t}) \\
        \robustnessFunc{\phi_1 \lor \phi_2}{t} &= \max( \robustnessFunc{\phi_1}{t}, \robustnessFunc{\phi_2}{t}) \\
        \robustnessFunc{\eventually{\phi}}{t} & = \max_{\tau \in [a,b]}(\robustnessFunc{\phi}{t+\tau}) \\
        \robustnessFunc{\always{\phi}}{t} &= \min_{\tau \in [a,b]}(\robustnessFunc{\phi}{t+\tau}) \\
        \robustnessFunc{\until{\phi_1}{\phi_2}}{t} &= \max_{\tau \in [a,b]}(\min(\robustnessFunc{\phi_2}{t+\tau}, \notag \\ &\qquad \min_{\tau^\prime \in [a, \tau]}\robustnessFunc{\phi_1}{t+\tau^\prime}))
    \end{align}
\end{subequations}
\subsection{STL Synthesis}
A typical synthesis problem is thus defined:
\begin{problem}[General Synthesis]\label{prob:general-synthesis}
    Given an environment $E \subseteq \mathbb{R}^n$, a set of dynamics constraints modeling a dynamical system $D: \mathbb{R}^n \to \mathbb{R}^n$ constraining a set of signals $\signalSet$, and an STL formula $\phi$, find a sequence of signal values that maximizes the robustness of $\phi$ while respecting $D$.
\end{problem}

Instead of a general synthesis problem, in this paper, we focus on a subclass of problems where the environment is represented as a transition system $TS = (G, \nodeLabelSet, \nodeLabelFunc)$. Where $G$ is a graph $(\nodeSet,\edgeSet, W)$, $\nodeSet$ is a set of nodes, $\edgeSet \subseteq \nodeSet \times \nodeSet$ is a set of edges, $W: \edgeSet \to \mathbb{Z}^+$ is a weight used to encode travel time, $\nodeLabelSet$ is a set of labels, and $\nodeLabelFunc: \nodeSet \to \mathcal{P}(\nodeLabelSet)$ is a labeling function which adds labels to nodes. In order to reason more effectively about large signal counts, we label the signals with a set of labels $\signalLabelSet$ and a labeling function $\signalLabelFunc: \signalSet \to \mathcal{P}(\signalLabelSet)$. These signals start at a node and propagate via edges. Our predicates then take the form $cnt_{\signalLabelElement, \nodeLabelElement} \geq \epsilon$, where robustness is semantically the number of signals with some label in a region with some label over the threshold $\epsilon$. Most MILP-based synthesis uses a more general set of dynamics, rather than this transition system abstraction~\cite {leahyScalableRobustAlgorithms2022}. However, as SMT needs more design for each dynamics encoding, we used this abstraction for this paper and leave adapting it to other types of dynamics as future work. Formally:
\begin{problem}[TS problem class]\label{prob:transition-system-synthesis}
    Given a transition system environment $TS$, a set of signals $\signalSet$, a set of signal labels $\signalLabelSet$, and an STL formula $\phi$, find a sequence of signal values that maximizes the robustness of the formula.
\end{problem}

\section{SMT encoding of STL robustness} \label{sec:smt_encoding}
To solve \Cref{prob:transition-system-synthesis}, we must encode the STL constraints as an SMT problem. This process is recursive, following the same procedure used for recursively encoding STL constraints as a MILP. In order to speed this up, we also memoize the encoding for a given formula at a given timestep so that, if it's encountered again, it doesn't need to reconstruct it. These encodings follow the form of~\Cref{eq:robustness_semantics} directly, which highlights the Boolean nature of SMT.
\subsection{Operator Encodings} \label{subsec:operator_encoding}
\begin{remark}[robustness encoding]
    In MILP STL literature~\cite{beltaFormalMethodsControl2019b,ramanModelPredictiveControl2014}, authors typically encode robustness and notate it using $r(\phi, t)$. They do this to show they are creating a new variable for that formula and time, which they then use in later constraints. We use $\robustnessFunc{\phi}{t}$ to indicate that we are \emph{not} creating a new variable, and instead just passing the constraint up to be included later. This difference is due to SMT preferring constraints that have structure that can be exploited for reasoning over, while MILP prefers lots of easier linear constraints relating variables.
\end{remark}

\subsubsection{Predicate} \label{subsubsec:predicate}
For a predicate comparing a signal $v \in \signalSet$ to a threshold $\epsilon$ we constrain the robustness to be:
\begin{equation}
    \label{eq:predicate}
    \robustnessFunc{v \text{ op } \epsilon}{ t} = \begin{cases}
        v - \epsilon & \text{op is } \ge \\
        \epsilon - v & \text{op is } \le \\
    \end{cases}
\end{equation}
In words, we set the robustness of a predicate to be the amount they exceed a constraint.
\subsubsection{Conjunction} \label{subsubsec:conjunction}
For a conjunction of two subformulas at time $t$, we constrain the robustness to be:
\begin{equation} \label{eq:conjunction}
    \robustnessFunc{\phi_1 \land \phi_2}{t}= \min_{i \in [1, 2]} (\robustnessFunc{\phi_i}{t})
\end{equation}
The robustness of $\phi_1$ and $\phi_2$ is the lower of the two, so we use \Cref{alg:min} to constrain it to be the lower of the two.

\subsubsection{Disjunction} \label{subsubsec:disjunction}
For a disjunction of two subformulas at time $t$, we constrain the robustness to be:
\begin{equation}
    \label{eq:disjunction}
    \robustnessFunc{\phi_1 \lor \phi_2}{ t} = \max_{i \in [1,2]}(\robustnessFunc{\phi_i}{ t)}
\end{equation}
The robustness of $\phi_1$ or $\phi_2$ is the greater of the two. \Cref{eq:disjunction} constrains it to be equal to the maximum using \Cref{alg:max}. 

\subsubsection{Negation} \label{subsubsec:negation}
For a negation of a subformula at time $t$, we constrain the robustness to be:

\begin{equation}
    \label{eq:negation}
    \robustnessFunc{\neg \phi}{t} = - \robustnessFunc{\phi}{t}
\end{equation}
Since negative robustness is how poorly it satisfies, the robustness of negation is just negating the robustness of the subformula.

\subsubsection{Eventually} \label{subsubsec:eventually}
For an eventually at time $t$, we constrain the robustness to be:

\begin{equation}
    \robustnessFunc{ \eventually{\phi}}{t} = \max_{\tau \in [a, \,b]} \robustnessFunc{\phi}{ t+\tau} \label{eq:eventually}
\end{equation}
Eventually is similar to saying OR for all the timesteps, so this constraint is very similar to \Cref{subsubsec:disjunction}. \Cref{eq:eventually} constrains the robustness for the formula to be equal to the largest of the timesteps within the range using \Cref{alg:max}
\subsubsection{Always} \label{subsubsec:always}
For an always at time $t$, we constrain the robustness to be:
\begin{equation}
    \robustnessFunc{\always{\phi}}{ t}  = \min_{ \tau \in [a, b]}(\robustnessFunc{\phi}{ t+\tau)}
\end{equation}
Always is similar to saying AND for all the timesteps, so this constraint is very similar to \Cref{subsubsec:conjunction}, using \Cref{alg:min} to constrain the robustness to be equal to the smallest robustness within the time range.

\subsubsection{Until} \label{subsubsec:until}
For an until at time $t$, we constrain the robustness to be:
\begin{equation}
\label{eq:until}
\begin{split}
    \robustnessFunc{ \until{\phi_1}{\phi_2}}{ t} = 
    \max_{\tau \in [a, b]} \Bigg( \min \Bigg( \robustnessFunc{\phi_2}{t+ \tau}, \\
    \min_{\tau^{\prime} \in [a, \tau]}\left( \robustnessFunc{\phi_1}{t+\tau^{\prime}} \right) \Bigg) \Bigg)
\end{split}
\end{equation}
where $\max$ is defined via \Cref{alg:max} and min is defined via \Cref{alg:min}

The outermost max is finding the maximally robust time point to release $\phi_1$. Each timepoint possibility is weighed by the equivalent of saying $(\always[t+a,\tau]{\phi_1})\land \phi_2$ or, in words, up until this timepoint $\phi_1$ has held, and at this time point $\phi_2$ holds.

\begin{algorithm}
\caption{SMT Max}
\label{alg:max}
\DontPrintSemicolon
\KwIn{$X$: List of real\_ref}
\KwOut{real\_ref that is the maximum of the input list}
$a \leftarrow \text{real\_ref}()$

$a \geq x \quad \forall x \in X$

$\bigvee_{x \in X} a = x$

\KwRet {$a$}

\end{algorithm}
\begin{algorithm}
\caption{SMT Min}
\label{alg:min}
\DontPrintSemicolon
\KwIn{$X$: List of real\_ref}
\KwOut{real\_ref that is the minimum of the input list}
$a \leftarrow \text{real\_ref}()$

$a \leq x \quad \forall x \in X$

$\bigvee_{x \in X} a = x$

\KwRet {$a$}

\end{algorithm}

\subsection{Transition System Dynamics Encodings} \label{subsec:ts_dynamics_encoding}
In addition to encoding the STL constraints for SMT, we must also add constraints for the dynamics. These will change based on the relevant formula and problem, as well as for different problem classes. The dynamics encoding diverges from the MILP solution \cite{leahyScalableRobustAlgorithms2022} in order to utilize our solvers better. The MILP solution makes this a numerical flow-balancing problem with many linear constraints, whereas our SMT dynamics approach uses a Boolean implication approach. 
 
Let $G = (\nodeSet, \edgeSet, W)$ be the transition-system graph, with node labels $\nodeLabelFunc:\nodeSet \to \mathcal{P}(\nodeLabelSet)$. Let $\nodeToInt: \nodeSet \to \{0,\dots,|\nodeSet|-1\}$ be the the node to integer bijection. We define a subset of nodes $\nodeSet_\nodeLabelElement \subseteq \nodeSet$ that have a given label $\forall \nodeLabelElement \in \nodeLabelSet$ with
\begin{equation} \label{eq:labeled_node_set}
    \nodeSet_\nodeLabelElement = \{u \in \nodeSet \mid \nodeLabelElement \in \nodeLabelFunc(u)\}
\end{equation}
We also define $\signalSet$ as a set of signals, with labels $\signalLabelFunc: \signalSet \to \mathcal{P}(\signalLabelSet)$, and for $\signalLabelElement \in \mathcal{P}(\signalLabelSet)$ similarly, we create a subset of signals $\signalSet_\signalLabelElement \subseteq \signalSet$ that have a given label $\forall \signalLabelElement \in \signalLabelSet$
\begin{equation} \label{eq:agent_set}
    \signalSet_\signalLabelElement = \{\signalElement \in \signalSet \mid \signalLabelElement \in \signalLabelFunc(\signalElement) \}
\end{equation}
As all of the temporal operators in STL have time bounds, the entire formula has a time bound. A time bound is the horizon over which the ability to determine satisfaction is guaranteed \cite{dokhanchi2014line}. Let $B \in \mathbb{N}$ be the time bound of the given formula, so time indices range over $\timeStep \in \{0, \dots, B\}$. Let $d: \nodeSet \times \nodeSet \to \mathbb{Z}_{\ge 0}$ be the all-pairs shortest path distance between nodes in the transition system from Bellman-Ford, weighted according to $W$.

Our specifications relate to labeled signal positions relative to labeled nodes. To reason about that, we need: \begin{inparaenum}
    \item a way to track individual signal positions; \item a way to roll those up into amounts of labeled signals at a given node; and \item a way to roll those up into amounts of labeled signals at any labeled node.
\end{inparaenum} Additionally, we need to restrict signal positions to follow the transition system, traversing via edges and taking time to do so.

To track signal positions over time, we introduce an integer SMT variable for every signal $\signalElement \in \signalSet$ and time $\timeStep \in \{0, \dots, B\}$ 
\begin{equation}\label{eq:ts_enc_position}
    \textrm{pos}_{\signalElement,\timeStep} \in \{-1\} \cup \{\nodeToInt(\nodeElement) | \nodeElement \in \nodeSet\}
    \end{equation}
where $-1$ encodes that an agent is in transit along an edge. To encode the initial placements $v_\signalElement^0 \in V$ we add the following constraint:
\begin{equation}\label{eq:ts_enc_initial_placements}
    \textrm{pos}_{\signalElement, 0} = \iota(v_\signalElement^0), \quad \forall s \in \signalSet \:.
\end{equation}
Reasoning about individual signal positions is less helpful than reasoning about node occupancy. We can define the occupancy count of a node $u \in \nodeSet$ by labeled signals $\forall \signalElement \in \signalSet_\signalLabelElement$ at time $k$:
\begin{equation} \label{eq:count_of_satisfying_signals}
    cnt_\signalLabelElement^k(u) = \sum_{\signalElement\in \signalSet_\signalLabelElement} [\textrm{pos}_{\signalElement,\timeStep} = \iota(u)]\:.
\end{equation}
Where $[\cdot]$ is the Iverson bracket defined as~\cite{knuthTwoNotesNotation1992}:
\begin{equation} \label{eq:iverson_bracket}
    [x] = \begin{cases}
    1 & x = \top \\
    0 & x = \bot
\end{cases}\:.
\end{equation}

When we follow \Cref{subsubsec:predicate} to encode predicates, we create state variables of the form $x_{\nodeLabelElement,\signalLabelElement,\timeStep}$ for relevant node and signal labels at a given time. For all of these variables created in the formula encoding process, we add constraints to make $x_{\nodeLabelElement,\signalLabelElement,\timeStep} = \min_{u\in \nodeSet_\nodeLabelElement} \mathrm{cnt}_\signalLabelElement^\timeStep(u)$ using \Cref{alg:min}. This sets that variable to be the amount of matching labeled signals at a labeled node with the least matching labeled signals, following the system used by \cite{leahyScalableRobustAlgorithms2022} for more equitable comparisons. We define a set of these variables $X = \{x_{\nodeLabelElement, \signalLabelElement, k} \mid  \forall \nodeLabelElement, \forall \signalLabelElement, \forall k \}$ for use constructing relevant nodes and signals to our formula. We also define a relation $R = (\nodeLabelElement, \signalLabelElement) \subseteq \nodeLabelSet \times \signalLabelSet$. This relation is constructed: 
\begin{equation}
    \label{eq:relevent_relation}
    \{(\nodeLabelElement, \signalLabelElement) \mid \exists k,  \exists x_{\nodeLabelElement, \signalLabelElement, k} \in X\} \:.
\end{equation}

Next we need to add propagation constraints, capturing how signals change over time according to the transition system. In order to simplify computation, we construct a smaller transition system using only nodes that contribute to the formula, connecting them with edges if there is a path in the overall transition system. We define these nodes of interest relative to each signal.
\begin{equation} \label{eq:nodes_of_interest}
    \begin{split}
            \textrm{NOI}_\signalElement = \{\nodeToInt(v_s^0)\} \cup \bigcup_{\signalLabelElement \in \signalLabelFunc(\signalElement)} \{\nodeToInt(\nodeElement) \mid \nodeElement \in \nodeSet, \\ \exists \nodeLabelElement \in \nodeLabelFunc(\nodeElement), \\ (\signalLabelElement, \nodeLabelElement) \in R \}
    \end{split}
\end{equation}
Then, for every $n \in \textrm{NOI}_\signalElement$ and every $\timeStep \in \{0, \dots, B-1\}$, we define the transition function, which uses position at time $k$ to imply which $\textrm{NOI}$ it can reach and force it in transit for that amount of time:
\begin{equation} \label{eq:transition_constraints}
\begin{split} 
        \mathrm{pos}_{\signalElement, \timeStep} = n \implies \bigvee_{\substack{m\in \textrm{NOI}_\signalElement \\ \timeStep+d(n,m) < B}} & \left( \bigwedge_{x=\timeStep+1}^{\timeStep+d(n,m) - 1}(\textrm{pos}_{\signalElement,x}=-1) \right. \\ & \quad \quad \left.\wedge \textrm{pos}_{\signalElement,\timeStep+d(n,m)}=m\right)
\end{split}    
\end{equation}

Finally, we can present the overall SMT problem to solve \Cref{prob:transition-system-synthesis}:
\begin{flalign}
\begin{aligned}\label{eq:full_SMT}
    \max\quad &\robustnessFunc{\phi}{0}\\
\text{s.t.}\quad & \text{dynamics constraints from (\ref{eq:ts_enc_position})-- (\ref{eq:transition_constraints})} \\    &\text{constraints }(\ref{eq:predicate})\text{--}(\ref{eq:until})
     \;\forall\text{ sub-formulae of }\phi.
\end{aligned}
\end{flalign}

\section{Simulations and Results}\label{sec:sim_and_results}

For ease of benchmarking, we utilized Capability Temporal Logic (CaTL)~\cite{leahyScalableRobustAlgorithms2022}, which is a fragment of STL with syntax given by:
\begin{equation}\label{CATL_syntax}
    \phi ::= T \mid \eventually{\phi} \mid \always{\phi}  \mid  \phi_1 \land \phi_2 \mid \phi_1 \lor \phi_2
\end{equation}
Where $T$ is a task defined as $\{d, L,  [\{\text{cap}, n\}]\}$ where $d$ is the number of time units the task must be completed for, $L \in \nodeLabelSet$ is a label from the environment, $\text{cap} \in \mathcal{P}(\signalLabelSet)$ is a set of capabilities required, and $n$ is the number of robots.
For our implementation, each labeled signal is an agent, and its labeled with the agents capabilities. 

We benchmarked our solver by creating a random environment $(\nodeSet, \edgeSet, W, \nodeLabelSet. \nodeLabelFunc)$, capabilities $\signalLabelSet$, and agent set $\signalSet$. We then generated a random CaTL formula. This combination of environment, capabilities, agents, and formula was then handed to each solver in order to compare state of the art MILP based solvers~\cite{leahyScalableRobustAlgorithms2022} with our SMT-based solver.

\subsection{Example Problem}\label{subsec:exampleprob}
\begin{figure}
    \centering
    \begin{subfigure}{0.48\linewidth}

\begin{tikzpicture}[scale=0.7]
  \definecolor{gridgreen}{RGB}{76,175,80}
  \definecolor{gridred}{RGB}{244,67,54}
  \definecolor{gridblue}{RGB}{33,150,243}
  \definecolor{gridyellow}{RGB}{255,255,0}
  
  \pgfmathsetmacro{\rows}{4}
  \pgfmathsetmacro{\cols}{5}

  \draw[fill={gridgreen}, draw=black, line width=0.5pt] (0, 0) rectangle (2, 1);
  \draw[fill={gridyellow}, draw=black, line width=0.5pt] (2, 0) rectangle (6, 1);
  \draw[fill={gridblue}, draw=black, line width=0.5pt] (0, 1) rectangle (1, 2);
  \draw[fill={gridred}, draw=black, line width=0.5pt] (0, 2) rectangle (1, 4);
    \draw[fill={gridgreen}, draw=black, line width=0.5pt] (0, 4) rectangle (6, 5);
  \draw[fill={gridyellow}, draw=black, line width=0.5pt] (1, 1) rectangle (2, 4);
  \draw[fill={gridred}, draw=black, line width=0.5pt] (2, 1) rectangle (4, 3);
  \draw[fill={gridblue}, draw=black, line width=0.5pt] (2, 3) rectangle (4, 4);
  \draw[fill={gridgreen}, draw=black, line width=0.5pt] (4, 1) rectangle (6, 2);
  \draw[fill={gridyellow}, draw=black, line width=0.5pt] (4, 2) rectangle (6, 4);
\end{tikzpicture} 
        \caption{Example environment, with different region labels $\nodeLabelSet = \{ \text{green}, \text{blue}, \text{yellow} \}$}
        \label{fig:environment}
    \end{subfigure}
    \hfill
    \begin{subfigure}{0.48\linewidth}
        \includegraphics[width=\textwidth]{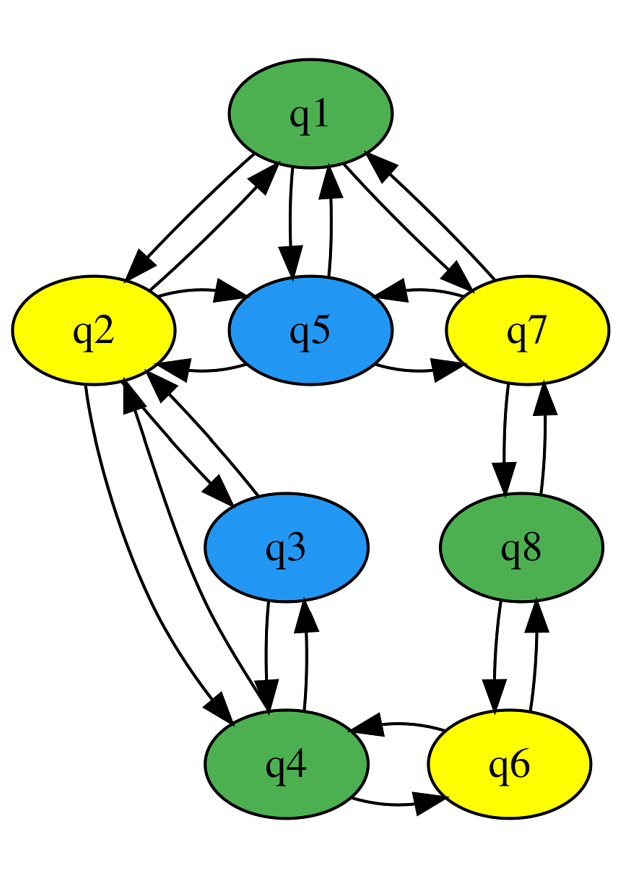}
        \caption{Transition system encoding of \Cref{fig:environment}; self-loops (staying in the region) and edge weights are not rendered for clarity of the diagram}
        \label{fig:transition-system-of-enviroment}
    \end{subfigure}%
    \setlength{\belowcaptionskip}{-10pt}
    \caption{Example environment and corresponding transition system. Obstacles (red regions) are omitted from the transition system.}
    \label{fig:Example-Environment}
\end{figure}

Using the environment shown in \Cref{fig:Example-Environment}, example capabilities $\signalLabelSet = \{ \text{Vis}, \text{Mo}, \text{IR} \}$ (camera, moisture sensors, and IR sensors, respectively), and an example agent set $S = \{ (1, q1, \{\text{Vis}\}), (2, q2, \{\text{Mo}, \text{IR}\}), (3, q4, \{\text{Vis}, \text{IR}\})\}$. Agents must satisfy a temporal logic specification, e.g., $\eventually[0,5]{\{2, \text{green}, [\{\text{IR}, 1\}]\}} \land \always[0,100]{\eventually[0,10]{\{3, \text{blue}, [\{\text{Vis}, 1 \}]\}}} $ This specification requires a robot with an IR sensor to spend 2 time units in a green region within the first 5 time steps, as well as a robot to spend at least 3 time units at a blue region every 10 time steps, for 100 time steps. The solvers then produce either a path for each robot or report the problem as infeasible.

\subsection{Results} \label{subsec:results}
All tests were run on an M1 Pro MacBook Pro with 16GB of RAM. All reported times are wall clock times. MILP formulations were solved with both Gurobi~\cite{gurobi} and SCIP~\cite{SCIPOptSuite10} via PuLP~\cite{dunningPuLPLinearProgramming2011}. SMT encodings were performed using Z3~\cite{demouraZ3EfficientSMT2008}

Gurobi was chosen because it is state-of-the-art in traditional MILP~\cite{yeMILPBenchLargescaleBenchmark2025}, but it is also a closed-source, uninspectable process with costly licensing. SCIP is a competitive open-source MILP solver~\cite{yeMILPBenchLargescaleBenchmark2025}, and therefore the most comparable option to the SMT solver we are using, Z3. 

\begin{figure}
    \centering
    \includegraphics[width=0.8\linewidth]{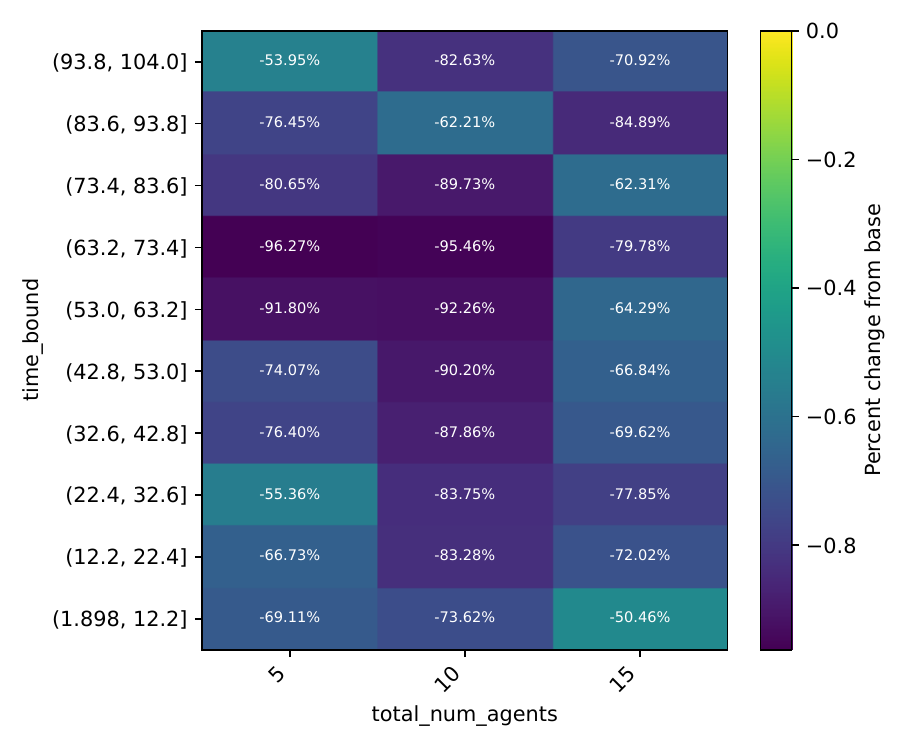}
    \caption{Percent change in solve time (lower is better) for parameter sweep of agent counts and time bounds for SMT compared to SCIP. SMT outperforms SCIP, with performance improving the most with lower agent counts and longer time horizons.}
    \label{fig:grid_scip}
\end{figure}

The results comparing with SCIP (\Cref{fig:grid_scip}) showcase the dramatic improvements that the SMT method has over MILP for this problem space, with results ranging from 50\% to 96\% faster depending on where in the parameter space the particular problem lies. 

\begin{figure}
    \centering
    \includegraphics[width=0.8\linewidth]{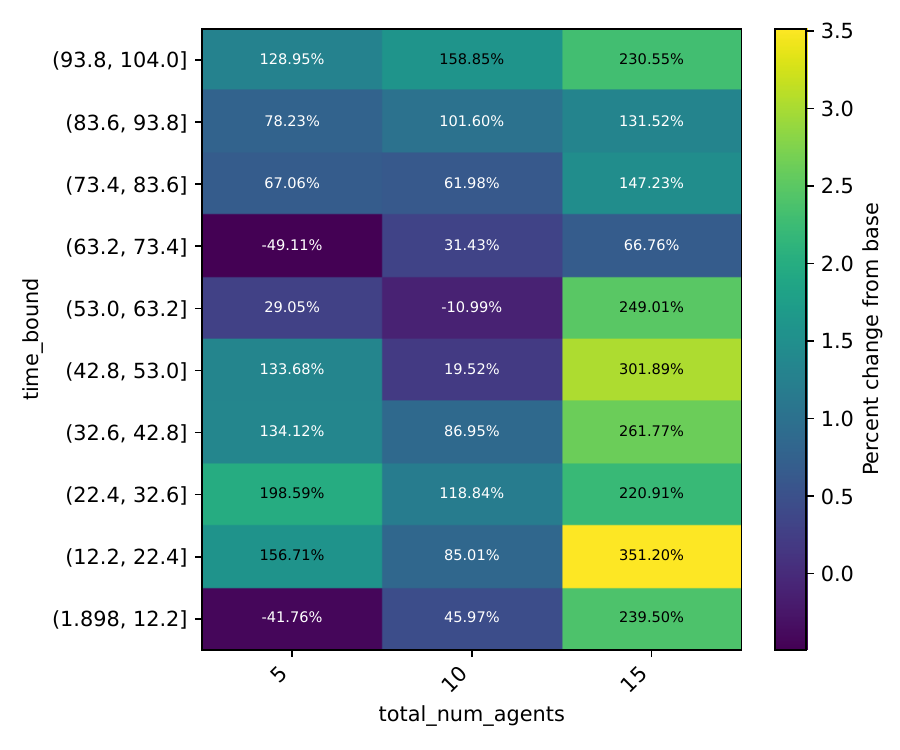}
    \caption{Percent change in solve time (lower is better) for a parameter sweep of agent counts and time bounds for SMT compared to Gurobi. Gurobi outperforms SMT, but SMT performance improves with lower agent counts and longer time horizons.}
    \label{fig:grid_gurobi}
\end{figure}

The comparison with Gurobi (\Cref{fig:grid_gurobi}) is best interpreted as a demonstration of the maturity of modern MILP solvers rather than as a head-to-head evaluation of comparable technologies. This observation also helps explain why MILP solvers have become the predominant backend in recent synthesis work. Although the SMT architecture is arguably better matched to the structure of this problem class, Gurobi's engineering and heuristics are sufficiently strong to offset that disadvantage. We include this comparison for completeness, and regard improving SMT heuristics and solver implementations to the point of being competitive with Gurobi as a direction for future work. 

\begin{figure}
    \centering
    \includegraphics[width=\linewidth]{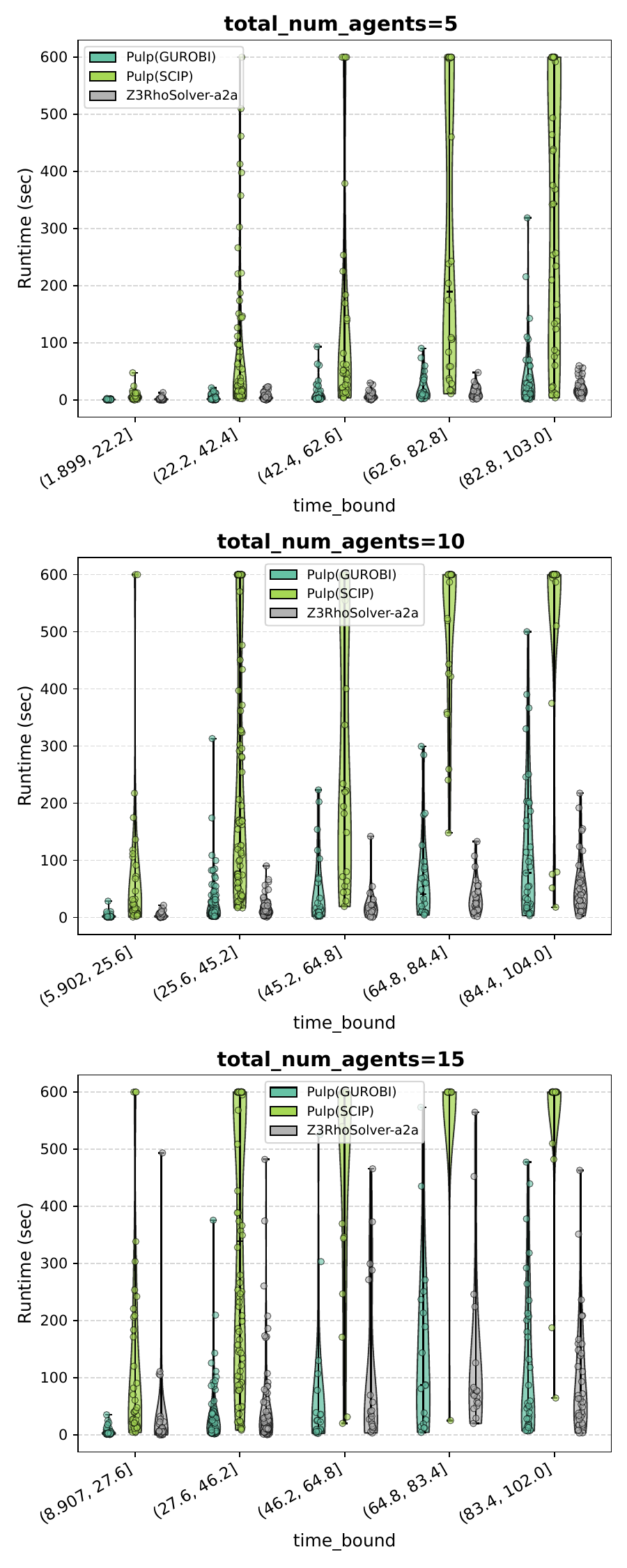}
    \caption{Distributions of solve times for agent counts and time bounds, showing that SMT performs the best at lower agent counts and long time horizons.}
    \label{fig:violin_plot_num_agents}
\end{figure}

Another included visualization of the results highlights the difference between SCIP and Gurobi solving times, with our SMT-based method sitting comfortably on the Gurobi side of the line (\Cref{fig:violin_plot_num_agents}).

Figures 2-4 consistently demonstrate an increase in SMT performance relative to MILP as time horizon increases. On the other hand, increases in the number of agents cause a decrease in SMT performance. These differences support our hypotheses about the differences between SMT and MILP performance on these problems. While MILP complexity grows with time horizon due to the increase of binary variables, SMT does not. Similarly, MILP formulations such as~\cite{leahyScalableRobustAlgorithms2022} encode agents into generic counts, allowing efficient computation of agent flow. Our SMT approach, however, treats agents as individuals; thus, the dynamics scale poorly relative to MILP.

\Cref{tab:statistics} displays solve counts for each approach. SMT consistently outperforms SCIP. SMT identifies the same number of infeasibilities at a faster computation rate, but it has more timeouts in the worst case.
\begin{table}
    \centering
    \begin{tabular}{|c|c|c|c|c|}
        \hline
        Solver & Solved& Feasible & Unfeasible & Timed out\\
        \hline
        Z3~\cite{demouraZ3EfficientSMT2008} & 679 & 138 & 541 & 41\\
        \hline
        Gurobi~\cite {gurobi} & 695 & 154 & 541 & 25\\
        \hline
        SCIP~\cite {SCIPOptSuite10} & 471 & 102 & 369 & 249\\
        \hline
    \end{tabular}
    \caption{Table showing counts for the runs graphed in Figures \ref{fig:grid_scip}, \ref{fig:grid_gurobi}, \ref{fig:violin_plot_num_agents}. Z3 is the SMT solver we used, and it consistently outperforms SCIP. It identifies the same number of infeasibilities as Gurobi, but has more timeouts in the worst case.}
    \label{tab:statistics}
\end{table}

\subsection{Replanning} \label{subsec:replanning}
\begin{figure}
    \centering
    \includegraphics[width=\linewidth]{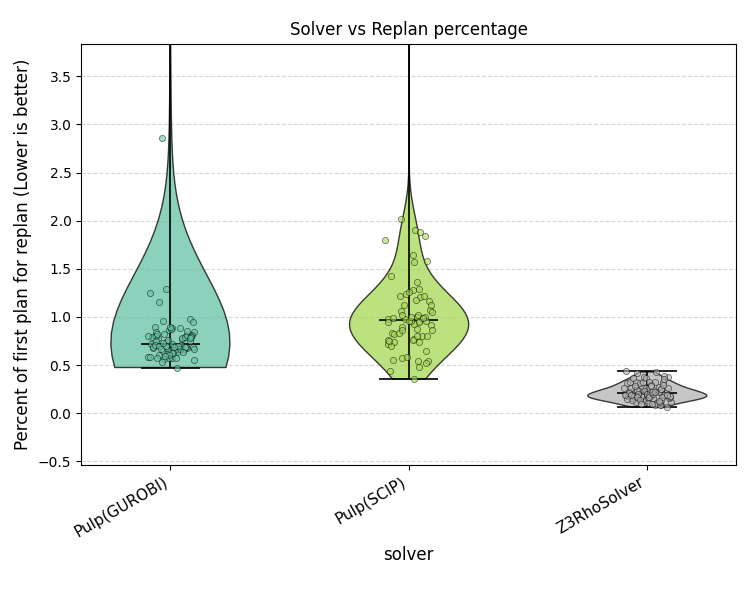}
    \caption{Percentage of first solve time taken to replan. This is zoomed in, as Gurobi and SCIP both have outliers so extreme that they hindered the usefulness of the graph.}
    \label{fig:replanning}
\end{figure}

Replanning using MILP is difficult due to the nature of the encoding, requiring most of the solution to be recomputed from scratch. SMT handles warm starts much more gracefully~\cite{dutertreFastLinearArithmeticSolver2006, aquinoReusingSolutionsModulo2021}. In order to test how the solvers performed at re-planning, we simulated discovering a non-traversable segment of the planned solution path by cutting one of the edges used in the solution.  \Cref{fig:replanning} shows replanning as a fraction of the initial planning time. SMT performs markedly better, with most of the experiments taking less than half as long to replan as the initial plan took to synthesize. 

\section{Conclusion and Future Work} \label{sec:conclusion}
SMT shows promise being used to synthesis control policies for temporal logic specifications, the combination of competitiveness on this type of problem (\Cref{subsec:results}) and better performance at replanning (\Cref{subsec:replanning}) makes SMT an attractive choice for synthesis of STL for some problems. 
Future work includes extending the encoding of dynamics beyond discrete transition systems to capture a wider variety of systems, creating encodings for other temporal logics besides STL, and adding other robustness requirements such as battery charge status or path length.

\printbibliography

\end{document}